\documentclass[aps,prl,twocolumn,showpacs]{revtex4}
\usepackage{amssymb}
\usepackage{amsmath, amsthm}

\begin{document}


\title{de Sitter space and the equivalence between $f(R)$ and 
scalar-tensor gravity}


\author{Valerio Faraoni}
\email[]{vfaraoni@ubishops.ca}
\affiliation{Physics Department, Bishop's University\\
Sherbrooke, Qu\`ebec, Canada J1M~0C8
}


\date{\today}

\begin{abstract}
It is shown that, when $f''\neq 0$, metric $f(R)$ gravity is 
completely equivalent to an $\omega=0 $ scalar-tensor theory 
with respect to perturbations of de Sitter space, contrary to 
previous expectations. Moreover, the stability 
conditions of de Sitter space with respect to homogeneous and 
inhomogeneous perturbations coincide in most scalar-tensor 
theories, as is the case in metric $f(R)$ gravity.
\end{abstract}


\maketitle


The acceleration of the cosmic expansion discovered using high 
redshift supernovae of type Ia \cite{SN} has been the subject of 
many attempts to understand its causes, and various models have 
been proposed for this phenomenon. As an alternative to dark 
energy 
models, modified gravity theories described by the 
action \footnote{We follow the notations of Ref.~\cite{Wald} 
and  $\kappa=8\pi G$, where $G$ is Newton's constant.}
\begin{equation}  
\label{action}
S=\frac{1}{2\kappa}\int d^4 x \sqrt{-g} \, f(R)+S^{(m)}    
\end{equation}
have been studied, where $f(R)$ is a non-linear function of the 
Ricci curvature $R$ that incorporates corrections to the 
Einstein-Hilbert action which is instead described by a linear 
function $f$. 
These modified gravity theories have been studied in the 
``metric formalism'', in which the action (\ref{action}) is 
varied with respect to the (inverse) metric $g^{ab}$, in the 
``Palatini 
formalism'', in which a Palatini variation with respect to both 
$g^{ab}$ and the (non-metric) connection $\Gamma^a_{bc}$ is 
performed \cite{Palatini}; and in the ``metric-affine'' context, 
in which also the matter part of the action $S^{(m)}$ is allowed 
to  depend on (and is varied with respect to) the connection 
$\Gamma^a_{bc}$ \cite{metricaffine}. Recently, it has been shown 
that most models proposed thus far in the metric 
formalism violate the weak-field constraints coming from Solar 
System experiments 
\cite{ChibaPLB03,ChibaSmithErickcek,Olmo, JinLiuLi}. 
However, there are still viable models \cite{Faulkneretal}. 
Earlier, and independent, interest in $f(R)$ gravity was 
motivated by scenarios of the early universe, such as 
Starobinsky inflation with 
$f(R)=R+aR^2-\Lambda$ and no scalar field \cite{Starobinsky}.

A common tool in the study of $f(R)$ gravity is its equivalence 
with a special scalar-tensor theory 
\cite{STequivalence,ChibaPLB03}. We 
restrict ourselves to 
the metric formalism in what follows. Then, one introduces the 
scalar field $\phi= R$ and the   
action~(\ref{action}) can be rewritten as
\begin{equation}  \label{STaction}
S=\frac{1}{2\kappa}\int d^4 x \sqrt{-g} \left[ \psi ( \phi) R 
-V(\phi) \right] +S^{(m)}     
\end{equation}
when $f''(R)\neq 0$, where a prime denotes differentiation with 
respect to $\phi$, 
\begin{equation}
\psi (\phi)\equiv f'( \phi) \;,
\end{equation}
and
\begin{equation}
V(\phi)=\phi f'(\phi)-f(\phi) \;.
\end{equation}
It is obvious that (\ref{action}) implies (\ref{STaction}) when 
$\phi=R$. Vice-versa, by varying (\ref{STaction}) with respect 
to 
$\phi$ one obtains
\begin{equation}
f''(R)\left( \phi-R\right)=0 \;,
\end{equation}
which yields $\phi=R$ if $f''(R)\neq 0$. The action 
(\ref{STaction}) describes a scalar-tensor theory with 
Brans-Dicke parameter $\omega=0$; similarly, Palatini $f(R)$ 
gravity can be reduced to a Brans-Dicke theory with parameter 
$\omega=-3/2$ \cite{STequivalence, ChibaPLB03}.

In a previous paper \cite{veto}, we raised doubts about the 
complete physical equivalence of the actions (\ref{action}) and 
(\ref{STaction}) (when $f''\neq0$), in the context of 
perturbations of de Sitter space. de Sitter solutions are 
important because  they are found very often to be late time 
attractors in the dynamics of the  accelerating  
universe, or inflationary attractors in the early universe,  
and because the weak-field limit of $f(R)$ gravity is obtained 
by expanding the relevant field equations around de Sitter space 
with a spherically symmetric perturbation 
\cite{ErickcekSmithKamionkowski, ChibaSmithErickcek,Olmo, 
JinLiuLi}.  In \cite{VFrapid,VFSN} we pointed out that the 
stability condition for de Sitter space with respect to 
{\em inhomogeneous} perturbations in $f(R)$ gravity, which we 
derived 
using a gauge-invariant formalism \cite{Bardeen,Hwang}, 
coincides with the corresponding stability condition with 
respect to {\em homogeneous} perturbations, and is
\begin{equation} \label{5}
\frac{ \left( f_0'\right)^2-2f_0 f_0''}{f_0' f_0''}\geq 0 \;.
\end{equation}
Here, and in the following, a zero subscript denotes quantities 
evaluated in the de Sitter background. 
As a consequence of this result, one can restrict oneself to 
the much simpler homogeneous perturbations of de Sitter 
space, which depend only on time and do not suffer from the 
notorious gauge-dependence problems. By contrast, it appears 
that in the scalar-tensor gravity theory (\ref{STaction}) the 
stability conditions with 
respect to homogeneous and inhomogeneous perturbations do not 
coincide \cite{veto, myPRD, VFrapid, VFSN}. The theory 
(\ref{STaction}) with $\omega=0$ is a very peculiar 
scalar-tensor theory, for which linear stability with respect to 
{\em inhomogeneous} 
perturbations corresponds again to
\begin{equation} \label{6}
\frac{ \left( f_0'\right)^2-2f_0 f_0''}{f_0' f_0''}\geq 0 
\end{equation}
as in $f(R)$ gravity, while stability with respect to 
{\em homogeneous} perturbations is equivalent  to
\begin{equation} \label{H7}
\frac{ \left( f_0'\right)^2-2f_0 f_0''}{f_0' }\geq 0  
\end{equation}
(beware of a typographical error in eq.~(33) of \cite{veto}).  
Here the subscript~$0$  denotes a quantity evaluated in the 
background de Sitter space $\left( H_0, \phi_0 \right)$.
Although the  difference between (\ref{6}) and (\ref{H7}) 
consists only of the 
factor $f_0''$ in the denominator and appears to be minimal, it 
is by no means trivial because one can not {\em a priori} decide 
that $f_0''$ should be positive. Based on this difference, 
doubts were raised in \cite{veto} on whether the two actions 
(\ref{action}) and (\ref{STaction}) are always physically 
equivalent. Complete physical equivalence was questioned also in 
\cite{Odintsov}. On 
the other hand, in the weak-field limit and in 
other studies of $f(R)$ gravity 
\cite{ChibaPLB03, ChibaSmithErickcek, Olmo, JinLiuLi}, it is 
found that this theory produces the same results as the 
$\omega=0$ scalar-tensor 
theory (\ref{STaction}). Here we  show how these two viewpoints 
can be reconciled in the light of recent results. The sign of 
$f''(R)$ {\em must} be positive: in Ref.~\cite{mattmodgrav} we 
studied the Dolgov-Kawasaki instability 
\cite{DolgovKawasaki} originally discovered  in the special 
model 
$f(R)=R-\mu^4/R$ (the prototype of corrections to the 
Einstein-Hilbert action aimed at explaining the present cosmic 
acceleration \cite{CCT, CDTT}). Since the field equations of the 
metric $f(R)$ formalism are the fourth order equations in the 
metric components
\begin{equation} \label{fieldeqs}
f'(R)R_{ab}-\frac{f(R)}{2}\, g_{ab}=\nabla_a\nabla_b 
f'(R)-g_{ab}\Box f'(R)+\kappa \, T_{ab} \;,
\end{equation}
by taking the trace one obtains the dynamical equation for the 
Ricci scalar 
\begin{equation} \label{trace}
3 f''(R)\Box R +3f'''(R)\nabla^c R\nabla_c 
R+f'(R)R-2f(R)=\kappa\, T \;.
\end{equation}
This shows that $R$ (or $\phi$ in the scalar-tensor description 
(\ref{STaction})) is a truly dynamical field, as opposed to the 
case of general relativity in which it satisfies the well-known 
algebraic equation $R=-\kappa\, T$ obtained from the trace of the 
Einstein equations. Therefore, in metric $f(R)$ gravity there 
is room for an instability in $R$ (which can also be seen as an 
instability in the matter sector \cite{DolgovKawasaki}). This 
instability  was discussed in \cite{mattmodgrav} for a  
general form of $f(R)$ and it was found that it is avoided 
if and 
only if $f''(R)\geq 0$ \footnote{See also a resolution of the 
instability in  special cases in \cite{SO1}, a discussion in 
\cite{SO2}, and 
stability issues in theories described by an action including 
the Gauss-Bonnet term in \cite{SO3}.}. This condition must be 
satisfied, in 
particular, for de Sitter spaces with constant curvature $R_0$  
and this is all is needed to show the equivalence of the 
inequalities (\ref{6}) and (\ref{H7}). Therefore, homogeneous 
and 
inhomogeneous perturbations of de Sitter space become equivalent 
both in metric $f(R)$ gravity and in its $\omega=0$ 
scalar-tensor formulation. The doubts raised in \cite{veto} 
about their equivalence are therefore dissipated.

Note that setting $f_0 ''>0$ is not 
justified {\em a priori}: the Ricci instability that plagues 
metric $f(R)$ gravity when $ f''(R)<0$ is by all means 
non-trivial. 
It arises because, contrary to the Einstein equations, 
eq.~(\ref{fieldeqs}) is of fourth order and its trace gives  
a dynamical equation for $R$. In the Palatini formalism, in which 
the field equations are only of second order, $R$ satisfies an 
algebraic equation as in general relativity, it is not a 
dynamical field, and there is no such 
instability \cite{Sotiriou}.

The stability condition $f''(R)\geq 0$ can be given a simple 
physical interpretation. Assume that the effective 
gravitational coupling $G_{eff}(R) \equiv G/f'(R)$ is positive; 
then, 
if $G_{eff}$ increases with the curvature, i.e.,
\begin{equation}
\frac{dG_{eff}}{dR}=\frac{-f''(R) G}{\left( f'(R) \right)^2}>0 
\;,
\end{equation}
at large curvature the effect of gravity becomes stronger, 
and since $R$ itself generates larger and larger curvature via 
eq.~(\ref{trace}), the effect of which becomes stronger and 
stronger because of an increased $G_{eff}(R)$, a positive 
feedback mechanism acts to destabilize the theory. If a small 
curvature grows and grows  without limit the system  runs 
away. If instead the 
effective gravitational coupling {\em decreases} when $R$ 
increases, which is achieved when $f''(R) >0$, a negative 
feedback mechanism operates which compensates for the increase 
in  $R$ and there is no running away of the solutions. It is 
curious that general  relativity, with $f''(R)=0$ and 
$G_{eff}=$~constant, is 
the  borderline case between stable behaviour ($f''>0$) and 
instability ($f''<0$).

At this point, we want to correct a mistake in the 
literature about homogeneous perturbations of de Sitter space in 
scalar-tensor gravity. It is stated in Ref.~\cite{SkeaBurd} that 
there are no stable de Sitter spaces in scalar-tensor gravity: 
this is clearly wrong as such examples abound in the literature 
(see, e.g., \cite{BarrowOttewill} and the references in 
\cite{mybook,FujiiMaeda}), as the condition (\ref{H7}) for 
stability shows. The error in the conclusion of \cite{SkeaBurd} 
seems to originate from incorrect signs in eqs.~(2.7)-(2.9) of 
that paper, which rule the evolution of homogeneous 
perturbations. In fact, de Sitter spaces can be stable with 
respect to more general inhomogeneous perturbations in more 
general gravity theories described by an action of the form 
\begin{equation}\label{H12}
S=\frac{1}{2\kappa}\int d^4 x\sqrt{-g} \left[ f\left( \phi, R 
\right) -\omega(\phi) g^{ab}\nabla_a\phi\nabla_b \phi -V(\phi) 
\right] 
\end{equation} 
which contains both $f(R)$ and scalar-tensor 
gravity as special cases. The gauge-invariant linear stability 
condition with respect to inhomogeneous perturbations is 
\cite{myPRD} 
\begin{equation} \label{H13}
\frac{     \frac{\partial^2 f}{\partial \phi^2 }\left.\right|_0 
-\frac{d^2 V}{d\phi^2}\left.\right|_0 
+ \frac{ R_0 f_{\phi R}^2 }{F_0} 
}{
\omega_0 \left( 1+\frac{ 3f_{\phi R}^2 }{ 2\omega_0 F_0} 
\right)
}  \leq 0 \;,
\end{equation}
where $F\equiv \partial f/\partial R$ and $f_{\phi R}\equiv 
\frac{ 
\partial^2 f}{\partial \phi\partial R}$.  Analogous stability 
conditions 
with respect to various kinds of classical and semiclassical 
instabilities were established in 
Refs.~\cite{BarrowOttewill,instabilities}. 

In our notations, the action of \cite{SkeaBurd} is
\begin{equation}\label{SkeaBurdaction}
S=\frac{1}{2\kappa} \int d^4x\sqrt{-g} \left[ \phi R 
-\frac{\omega(\phi)}{\phi} \, g^{ab} \nabla_a \phi\nabla_b \phi 
-V(\phi) \right] \;.
\end{equation}
In the spatially flat Friedmann-Lemaitre-Robertson-Walker metric 
$ds^2=-dt^2 +a^2(t)\left( dx^2+dy^2+dz^2 \right) $, the field 
equations assume the form 
\begin{widetext}
\begin{eqnarray}
&& H^2 = \frac{ \omega(\phi)}{6} \left( \frac{\dot{\phi}}{\phi} 
\right)^2 +\frac{V(\phi)}{6\phi}-\frac{H\dot{\phi} }{\phi} 
\;,\\
&&\label{Hamconstraint} \nonumber \\
&& \dot{H}= -\frac{\omega(\phi)}{2} \left( \frac{ 
\dot{\phi}}{\phi} \right)^2 +2H \, \frac{\dot{\phi}}{ \phi} 
+\frac{1}{2\left( 2\omega+2 \right) \phi}\left[ 
\frac{d\omega}{d\phi}\, \dot{\phi}^2 +\phi\, \frac{dV}{d\phi} 
-2V \right] \;, \label{H2} \\
&&\\
&& \ddot{\phi}+\left( 3H +\frac{\dot{\omega}}{2\omega+3}\right) 
\dot{\phi} =\frac{1}{2\omega+3} \left( -\phi \, \frac{dV}{d\phi} 
+2V(\phi) \right) \;.
\end{eqnarray}
\end{widetext}

A de Sitter space $\left( H_0, \phi_0 \right)$ corresponds to
\begin{equation}\label{H18}
H_0^2=\frac{V_0}{6\phi_0} \;, \;\;\;\;\;\;\;
\phi_0 V_0'=2V_0=12H_0^2 \phi_0 =R_0\phi_0 \;.
\end{equation}
Homogeneous perturbations of this solution are given by 
$ 
H(t)=H_0+\delta H(t)$ and $\delta \phi(t)=\phi_0+\delta 
\phi(t)$; then, the Hamiltonian constraint 
(\ref{Hamconstraint}) yields, to linear order,
\begin{equation} \label{H19}
\delta H= \left[ \frac{V_0'}{12H_0 \phi_0 } -\frac{V_0}{12H_0 
\phi_0^2} \right] \delta \phi -\frac{1}{2\phi_0} \delta 
\dot{\phi} \;.
\end{equation}
Eq.~(\ref{H2}) yields, to first order, 
\begin{equation}\label{H20}
\delta \dot{H}=\frac{2H_0}{\phi_0}\, \delta \dot{\phi}+\frac{
\left( \phi_0 V_0''-2V_0' +2V_0/\phi_0\right)}{2\phi_0 \left( 
2\omega_0 +3 \right)}\, \delta \phi \;.
\end{equation}
By substituting eq.~(\ref{H20}) into eq.~(\ref{H19}), these two 
equations decouple and one obtains the wave equation for the 
linear scalar field perturbation $\delta \phi$
\begin{equation}
\delta \ddot{\phi} +3H_0 \, \delta \dot{\phi} + 
\frac{
\left( \phi_0 V_0''-2V_0' +2V_0/\phi_0\right)}{2\phi_0 \left( 
2\omega_0 +3 \right)}\,\delta \phi=0 \;.
\end{equation}
The coefficient of $\delta \phi$ in the last term on the left 
hand side plays the role of a mass squared and stability 
corresponds to this quantity being non-negative. By using 
eq.~(\ref{H18}), the stability condition becomes
\begin{equation}
\frac{ \phi_0 V_0'' -12 H_0^2 }{ 2\omega_0 +3  } \geq 0 \;.
\end{equation}
It is straightforward to show that this inequality coincides 
with the stability condition (\ref{H13}) by using $ f\left( 
\phi, R \right)=\phi R$, $R_0=12H_0^2$, $F=\phi$, $f_{\phi 
R}=1$, and replacing $\omega_0$ with $ \omega_0/\phi_0$ in the 
denominator of (\ref{H13}) to account for the different form of 
the coefficient of the kinetic term of $\phi$ in the actions 
(\ref{H12}) and (\ref{SkeaBurdaction}). There are, of course, de 
Sitter spaces which are stable with respect to homogeneous 
perturbations. 

Now, a general scalar-tensor theory of the form
\begin{equation}\label{twostars}
S=\frac{1}{2\kappa}\int d^4 x \sqrt{-g} \left[ \psi(\phi) R 
-\frac{\omega(\phi)}{\phi} \, g^{ab} \nabla_a \phi\nabla_b\phi 
-V(\phi) \right]
\end{equation}
can be recast in the form (\ref{SkeaBurdaction}) by introducing 
the new scalar field $\varphi (\phi) \equiv \psi(\phi)$, 
obtaining
\begin{equation}\label{star}
S=\frac{1}{2\kappa}\int d^4 x \sqrt{-g} \left[ \varphi R 
-\frac{\bar{\bar{\omega}}(\varphi)}{\varphi} \, g^{ab} \nabla_a 
\varphi\nabla_b\varphi 
-U(\varphi) \right] 
\end{equation}
where
\begin{eqnarray}
\bar{\bar{\omega}}(\varphi) &=& \frac{ \varphi \, \omega\left[ 
\psi^{-1}(\varphi) \right]}{\psi^{-1}(\varphi) } \left( 
\frac{d\psi}{d\phi} \right)^{-2} \;,  \\
&&\nonumber \\
U( \varphi)&=& V\left[ \psi^{-1}(\varphi)\right] \;.
\end{eqnarray}
The action (\ref{star}) coincides with the action 
(\ref{SkeaBurdaction}) that we have just studied. 
(\ref{SkeaBurdaction}) is equivalent to (\ref{twostars}) 
provided that the function $\varphi=\psi(\phi) $ is invertible 
and the derivatives of $\psi$ and its inverse $\psi^{-1}$ are 
well-defined. Under this assumption (which is not always 
satisfied for the choices of $\psi(\phi)$ found in the 
literature --- 
see, e.g., \cite{LiddleWandsTorresVucetich}), we have shown 
above that {\em the stability conditions of de Sitter space with 
respect to homogeneous and inhomogeneous perturbations 
coincide}. Therefore, one can restrict oneself to considering 
the much simpler homogeneous perturbations.

\begin{acknowledgments}

The author thanks T.P. Sotiriou, J.C. Campos de Souza and M.N. 
Jensen for useful discussions. This work was supported by a 
Bishop's University Research Grant and by the Natural  Sciences 
and Engineering  Research Council of Canada.
\end{acknowledgments}


\begin{thebibliography}{99}


\bibitem{Wald} R.M. Wald, {\em General Relativity} (Chicago 
University Press, Chicago, 1984).


\bibitem{SN} A.G. Riess {\em et al.}, {\em Astron. J.} 
{\bf 116}, 1009 (1998); {\em Astron. J.} {\bf 118}, 2668 
(1999);
{\em Astrophys. J.} {\bf 560}, 49 (2001);
{\em Astrophys. J.} {\bf 607}, 665 (2004);
S. Perlmutter {\em et al.}, {\em Nature} {\bf 391}, 51 (1998);
{\em Astrophys. J.} {\bf 517}, 565 (1999); 
J.L. Tonry {\em et al.}, {\em Astrophys. J.} {\bf 594}, 
1 (2003);
R.  Knop {\em et al.}, {\em Astrophys. J.} {\bf 598}, 102 
(2003);
B. Barris {\em et al.}, {\em Astrophys. J.} {\bf 602}, 571 
(2004);
A.G. Riess {\em et al.}, astro-ph/0611572.

\bibitem{Palatini} 
D.N. Vollick, {\em Phys. Rev. D} {\bf 68}, 
063510 (2003); {\em Class. Quant. Grav.} {\bf 21}, 3813 (2004);
X.H. Meng and P. Wang, {\em Class. Quant. 
Grav.} {\bf 20}, 4949 (2003);  {\em Class. Quant. Grav.} {\bf 
21}, 951 (2004); {\em Phys. Lett.} {\bf 584B}, 1 (2004);
E.E. Flanagan, {\em Phys. Rev. Lett.} {\bf 
92}, 071101 (2004);
T. Koivisto, {\em Phys. Rev. D} {\bf 73}, 083517 (2006); 
{\em Class. Quant. Grav.} {\bf 23}, 4289 (2006); 
T. Koivisto and H. Kurki-Suonio, {\em Class. Quant. Grav.} {\bf 
23}, 2355 (2006); 
P. Wang, G.M. Kremer, D.S.M. Alves, and X.H. Meng, 
{\em Gen. Rel. Grav.} {\bf 38}, 517 (2006);
G. Allemandi, M. Capone, S. Capozziello, and M. Francaviglia, 
{\em Gen. Rel. Grav.} {\bf 38}, 33 (2006);
S. Nojiri and S.D. Odintsov, hep-th/0601213.

\bibitem{metricaffine} T.P. Sotiriou, {\em Class. Quant. 
Grav.} {\bf 23}, 5117 (2006);
N.J. Poplawski, {\em Class. Quant. Grav.} {\bf 23}, 2011 
(2006); {\bf 23}, 4819 (2006).

\bibitem{ChibaPLB03} T. Chiba, {\em Phys. Lett. B} {\bf 575}, 
1 (2003).

\bibitem{ChibaSmithErickcek} T. Chiba, T. 
Smith, and A. Erickcek,  astro-ph/0611867.

\bibitem{Olmo} G.J. Olmo, to appear in 
{\em Phys. Rev. D} (gr-qc/0612047).

\bibitem{JinLiuLi} X.-H. Jin, D.-J Liu, and X.-Z Li, 
astro-ph/0610854.

\bibitem{Faulkneretal} T. Faulkner, M. Tegmark, E.F. Bunn, and 
Y. Mao, astro-ph/0612569; S. Nojiri and S.D. Odintsov, 
hep-th/0611071.

\bibitem{Starobinsky} A.A. Starobinsky, {\em Phys. Lett. B} {\bf 
91}, 99 (1980).

\bibitem{STequivalence} P.W. Higgs 1959, {\em Nuovo Cimento} 
{\bf 11}, 816 (1959);   P. Teyssandier and P. Tourrenc, 
{\em J. Math. Phys.} {\bf 24}, 2793 (1983); 
B. Whitt, {\em Phys. Lett. B} {\bf 145}, 
176 (1984); D. Wands, {\em 
Class. Quant. Grav.} {\bf 11}, 269 (1994).

\bibitem{veto} V. Faraoni, {\em Phys. Rev. D} {\bf 74}, 023529 
(2006); erratum: {\bf 75}, 029902(E) (2007).

\bibitem{ErickcekSmithKamionkowski} A. Erickcek, T.L. Smith, 
and M. Kamionkowski, {\em Phys. Rev. D} {\bf 74}, 
121501(R) (2006).

\bibitem{VFrapid} V. Faraoni, {\em Phys. Rev. D} {\bf 72}, 
061501(R) (2005).
\bibitem{VFSN} V. Faraoni and S. Nadeau, {\em Phys. 
Rev. D} {\bf 72}, 124005 (2005).

\bibitem{Bardeen} J.M. Bardeen, {\em Phys. Rev. D} {\bf 22},  
1882 (1980);
G.F.R. Ellis and M. Bruni, {\em Phys. Rev. D} {\bf 40}, 1804 
(1989); 
G.F.R. Ellis, J.-C. Hwang and M. Bruni, {\em Phys. Rev. D} 
{\bf 40},  1819 (1989); 
G.F.R. Ellis, M. Bruni and J.-C. Hwang, {\em Phys. Rev. D} 
{\bf 42}, 1035 (1990).

\bibitem{Hwang} J.-C. Hwang, {\em Class. Quant. Grav.}  {\bf 
7},  1613 (1990);
{\bf 14},  1981 (1997); 3327; {\bf 15},  1401 (1998);  1387; 
{\em Phys. Rev. D} {\bf 42},  2601 (1990); {\bf 53},  762 (1996); 
J.--C. Hwang and H. Noh, {\em Phys. Rev. D} {\bf 54}, 1460 (1996).

\bibitem{myPRD} V. Faraoni, {\em Phys. Rev. D} {\bf  
70},  044037 (2004).

\bibitem{mattmodgrav} V. Faraoni, {\em Phys. Rev. D} {\bf 
74}, 104017 (2006).

\bibitem{DolgovKawasaki} A.D. Dolgov and M. Kawasaki, {\em Phys. 
Lett. B} {\bf 573}, 1 (2003).

\bibitem{CCT} S. Capozziello, S. Carloni, and A. Troisi, 
astro-ph/0303041.

\bibitem{CDTT} S.M. Carroll, V. Duvvuri, M. Trodden, and M.S. 
Turner, {\em Phys. Rev. D} {\bf 70}, 043528 (2004).

\bibitem{Sotiriou} T.P. Sotiriou, {\em Phys. Lett. B} {\bf 645}, 
389 (2007).

\bibitem{SkeaBurd} J.E.F. Skea and A.B. Burd, {\em Phys. Lett. 
B} {\bf 232}, 452 (1989).

\bibitem{BarrowOttewill} J.D. Barrow and A.C. Ottewill, {\em 
J. Phys. A} {\bf 16}, 2757 (1983).

\bibitem{mybook} V. Faraoni, {\em Cosmology in Scalar-Tensor 
Gravity} (Kluwer Academic, Dordrecht, 2004).

\bibitem{FujiiMaeda} Y. Fujii and K. Maeda, {\em The 
Scalar-Tensor Theory of Gravity} (CUP, Cambridge, 2003).

\bibitem{instabilities} I. Navarro and K. Van 
Acoleyen, {\em J. Cosmol. Astropart. Phys.} {\bf 03}, 008 
(2006); 
G. Cognola, E. Elizalde, S. Nojiri, S.D. 
Odintsov, and S. Zerbini, {\em J. Cosmol. Astropart. Phys.} {\bf 
02}, 010 (2005); {\em J. Phys. A} {\bf 39}, 6245 (2006); {\em J. 
Phys. A} {\bf 39}, 6245 (2006); 
T. Clifton and J.D. Barrow, gr-qc/0511036; {\em Class. Quant. 
Grav.} {\bf 23}, L1 (2006); {\em Phys. Rev. D} {\bf 72}, 103005 
(2005);
A. N\'{u}nez and S. Solganik, {\em Phys. Lett. B} 
{\bf 608}, 189 (2005); hep-th/0403159;
T. Chiba, {\em J. Cosmol. Astropart.  Phys.} {\bf 0503}, 008  
(2005);
P. Wang, {\em Phys. Rev. D} {\bf 72}, 024030 (2005);
J.D. Barrow and S. Hervik, {\em 
Phys. Rev. D} {\bf 73}, 023007 (2006);
A. De Felice, M. Hindmarsh, and M. Trodden, astro-ph/0604154;
G. Calcagni, B. de Carlos, and A. De Felice, 
{\em Nucl. Phys. B} {\bf 752}, 404 (2006);
A. Dolgov and D.N. Pelliccia, 
{\em Nucl. Phys. B} {\bf 734}, 208 (2006);
J. Traschen and C.T. Hill, {\em Phys. Rev. D} {\bf 33}, 3519 
(1986); 
V. M\"{u}ller, 
H.-J. Schmidt, and A.A. Starobinsky, {\em Phys. Lett.} {\bf 
202B}, 198 (1988); H.-J. Schmidt, {\em Class. Quant. Grav.} {\bf 
5}, 233 (1988); A. Battaglia Mayer and H.-J. Schmidt, {\em Class. 
Quant. Grav.} {\bf 10}, 2441 (1993);
O. Bertolami, {\em Phys. Lett. B} {\bf 186}, 161 
(1987); M.R. Setare, {\em Phys. Lett. B} {\bf 644}, 99 (2007).

\bibitem{LiddleWandsTorresVucetich}  A.R. Liddle and D. Wands, 
{\em Phys. Rev. D} {\bf 45}, 2665 
(1992);
D.F. Torres and H. Vucetich, {\em Phys. Rev. D} {\bf 54}, 7373 
(1996). 

\bibitem{Odintsov} S. Capozziello, S. Nojiri, S.D. Odintsov, and 
A. Troisi, {\em Phys. Lett. B} {\bf 639}, 135 (2006).

\bibitem{SO1} S. Nojiri and S.D. Odintsov, {\em Phys. Rev. D} 
{\bf 68}, 123512 (2003); {\em Gen. Rel. Grav.} {\bf 36}, 1765 
(2004); hep-th/0608008.

\bibitem{SO2} S. Capozziello, S. Nojiri, and S.D. Odintsov, {\em 
Phys. Lett. B} {\bf 634}, 93 (2006).

\bibitem{SO3} G. Cognola, E. Elizalde, S. Nojiri, S.D. 
Odintsov, and S. Zerbini, hep-th/0611198; {\em Phys. Rev. D} 
{\bf 73}, 084007  (2006).

\end{thebibliography}

\end{document}